\documentclass[12pt,preprint]{aastex}
\begin{document}

\title{The Effects of Planets and Brown Dwarfs on Stellar
Rotation and Mass-Loss}
\author{Mario Livio}
\affil{Space Telescope Science Institute\\
3700 San Martin Drive\\Baltimore, MD 21218\\
mlivio@stsci.edu}
\and
\author{Noam Soker}
\affil{Department of Astronomy\\
University of Virginia\\and\\Department of Physics\\
University of Haifa at Oranim\\Oranim, Tivon 36006, Israel\\
soker@physics.technion.ac.il}

\begin{abstract}
We examine the effects of the engulfment of planets by giant stars on the evolution of late-type stars.  We show that the rate at which dynamo-generated magnetic energy is being released exceeds 10\% of the wind kinetic energy when the orbital angular momentum of the engulfed planet is more than ten times the angular momentum of the star as it leaves the main sequence.  A significant enhancement in the mass-loss rate may be expected in this case, due to the formation of cool magnetic spots. We use the existing sample of extrasolar planets to estimate that at least 3.5\% of the evolved solar-type stars will be significantly affected by the presence of planetary companions.
\end{abstract}
\keywords{planetary systems -- stars: evolution -- stars: mass-loss -- stars: magnetic fields -- stars: statistics -- stars: low mass, brown dwarfs}

\section{Introduction}

Radial velocity surveys of nearby stars have been enormously
successful in discovering extrasolar planets. These surveys have
demonstrated that a significant fraction of the stars in the solar
neighborhood have massive planets with orbital radii that are
substantially smaller than Jupiter's 
\citep[e.g.][]{marcy00,udry01,butler01,vogt02}.
The engulfing of these planets by their parent stars is an inevitable
outcome of stellar evolution (either directly or due to tidal interaction)
in the red giant branch (RGB) or asymptotic giant branch (AGB) phase
\citep[e.g.][]{livio84,siess99a,siess99b}.
The capture of planetary-mass companions has been suggested
to be the potential cause for the high rotational velocities
($V_{\mathrm{rot}}\sin i\ga10$~km~s$^{-1}$) observed in some
field red giants 
\citep{stefanik01}
and blue horizontal
branch stars 
\citep[e.g.][]{peterson83,soker00}.

In the present work we examine specifically the effects of
planet/brown dwarf ``swallowing'' on the rotation rates of and
mass loss from giants. In particular, we estimate the expected
statistics of influenced stars.

\section{Rotation and Mass Loss in Evolved Late-Type Stars}

When a star in the RGB or AGB phase engulfs a planet/brown dwarf the
latter deposits its angular momentum into the giant's envelope.
 The resulting rotation frequency, as a fraction of the critical
 stellar rotation frequency $\omega_k\equiv(GM_*/R^3_*)^{\frac{1}{2}}$
 (where $M_*$, $R_*$ are the stellar mass and radius respectively)
 is given by
\begin{equation}
\frac{\omega}{\omega_k}\simeq0.10
\left( \frac{M_c}{0.01~\mathrm{M}_{\odot}} \right)
\left( \frac{M_\mathrm{env}}{\mathrm{M}_{\odot}} \right)^{-1}
\left( \frac{k^2_g}{0.1} \right)^{-1} 
\left( \frac{a}{R_*} \right)^{\frac{1}{2}}~~.
\end{equation}
Here $M_c$ is the companion mass, $M_{\mathrm{env}}$ is the mass of the
giant's envelope, $Mk^2_gR^2$ is the star's moment of inertia and $a$ is the initial
separation between the giant and the companion. Since stars like the
Sun rotate only at a fraction of a percent of their critical rate
(on the main sequence) and this fraction is further reduced
(by conservation of angular momentum) to
\begin{equation}
\left( \frac{\omega}{\omega_k} \right)_{\mathrm{AGB}} \lesssim0.1
\left( \frac{\omega}{\omega_k} \right)_{\mathrm{MS}}
\left( \frac{R_{\mathrm{MS}}}{0.01R_{\mathrm{AGB}}} \right)^{\frac{1}{2}}~~,
\end{equation}
as the stars ascend the RGB or AGB, we see from eq.~(1) that even
Jupiter-mass planets can have a very significant effect on the
rotation rate of late-type stars. Even when the increase in central condensation is taken into account, typically, $(\omega /\omega_k)_\mathrm{AGB} \sim(0.01-0.1)(\omega /\omega_k)_\mathrm{MS}$ \citep[e.g.][]{eriguchi92}.

The effects on mass loss rates are more difficult to estimate because
they involve the processes of generation of magnetic activity
(especially in spots) and mass loss via radiation pressure on
molecules and dust, both of which are rather poorly understood.
Nevertheless, it has been shown semi-quantitatively that the spiraling-in
process of the companion inside the giant's envelope can result in a
dynamo generation of a magnetic field 
\citep{regos95}.

Here we shall adopt a simplified, more heuristic approach 
\citep[similar to that of][]{sokerharpaz92}
to estimate at what rotation rates we can expect significant 
effects on mass loss. Dynamo-generated magnetic fields are limited by the 
requirement that the amplification timescale $\tau_a$ be of the order of 
the buoyant rise time of magnetic flux tubes 
\citep[e.g.][\S8.7]{parker79}.
In a typical stellar $\alpha\omega$ dynamo the amplification timescale
is given approximately by 
\citep[e.g.][]{zeldovich83}
 \begin{equation}
\tau_a\simeq\pi\left( \frac{2}{\alpha R_* k\nabla\omega} \right)^{\frac{1}{2}}~~,
\end{equation}
where $\alpha=\ell\omega_o/3$ is determined by convection ($\ell$ is
the mixing length and $\omega_o$ the surface angular velocity),
$k$ is the wave number and $\nabla\omega$ is a measure of differential
rotation. The critical wave number $k_c$ below which dynamo waves are
amplified is given by ($\nu_{\tau}$ is the turbulent viscosity)
\begin{equation}
k^3_c\simeq\alpha R_* \nabla\omega/2\nu^2_{\tau}~~.
\end{equation}
Adopting the same values as Zeldovich et~al.\ (1983), 
$\nabla\omega=\omega_o/(0.3R_*)$, $\nu_{\tau}=\ell v_c/3$, where $v_c$ is the convective velocity, and taking the wave number to be equal to the critical one, we find
\begin{equation}
\tau_a\simeq\frac{2\pi}{\omega_o}
\left( \frac{v_c}{7\ell\omega_o} \right)^{\frac{1}{3}}~~.
\end{equation}
For typical AGB star parameters, $v_c\simeq10$~km~s$^{-1}$,
$<$$\ell$$>\sim R_*/4$, the obtained average amplification time is of the
order of $\tau_a\simeq9\tau_{\mathrm{rot}}$,
 where $\tau_{\mathrm{rot}}$ is the rotation period.
 The local rise time of a flux tube is given approximately by 
\citep{parker79}
\begin{equation}
\tau_r\simeq\frac{\Lambda v_c}{V^2_A}=\frac{4\pi\rho\Lambda v_c}{B^2}~~,
\end{equation}
where $V_A$ is the Alfven speed, $B$ is the magnetic field strength,
$\Lambda\sim r/2$ \citep[where $r$ is the radial distance from the center; e.g.][]{sokerharpaz99} is the local scaleheight and $\rho$ is the density.
The density in the convective zone of an AGB star can be well
approximated by 
\citep{soker92}
\begin{equation}
\rho(r)\simeq2\times10^{-8}
\left( \frac{M_{\mathrm{env}}}{0.1~\mathrm{M}_{\odot}} \right)
\left( \frac{R_*}{300~\mathrm{R}_{\odot}} \right)^{-1}
\left( \frac{r}{100~\mathrm{R}_{\odot}} \right)^{-2}\ \mathrm{g~cm}^{-3}~~.
\end{equation}
Equating the flux tube rise time $\tau_r$ to the regeneration and amplification time
$\tau_a$ yields a value for the magnetic field
\begin{equation}
B\simeq1.8
\left( \frac{M_*}{\mathrm{M}_{\odot}} \right)^{\frac{1}{3}}
\left( \frac{R_*}{300~\mathrm{R}_{\odot}} \right)^{-\frac{4}{3}}
\left( \frac{M_{\mathrm{env}}}{0.1~\mathrm{M}_{\odot}} \right)^{\frac{1}{2}}
\left( \frac{v_c}{10~\mathrm{km~s}^{-1}} \right)^{\frac{1}{3}}
\left( \frac{\omega_o/\omega_k}{10^{-3}} \right)^{\frac{2}{3}}
\left( \frac{r}{100~\mathrm{R}_{\odot}} \right)^{-\frac{1}{2}}\ \mathrm{G}~~.
\end{equation}
Magnetic energy is thus being released at a rate
\begin{eqnarray}
\dot{E}_B&=&\frac{1}{\tau_a}\int\frac{B^2}{8\pi} 4\pi r^2 d r\nonumber\\ 
&\simeq&10^{28}
\left( \frac{M_{\mathrm{env}}}{0.1~\mathrm{M}_{\odot}} \right)
\left( \frac{v_c}{10~\mathrm{km~s}^{-1}} \right)^{\frac{1}{3}}
\left( \frac{R_*}{300~\mathrm{R}_{\odot}} \right)^{-\frac{7}{3}}
\left( \frac{M_*}{\mathrm{M}_{\odot}} \right)^{\frac{4}{3}}
\left( \frac{\omega_o/\omega_k}{10^{-3}} \right)^{\frac{8}{3}}~\mathrm{erg~s}^{-1}~~.
\end{eqnarray}

The kinetic energy associated with the mass loss from an AGB star is of
order
\begin{equation}
\dot{E}_{\mathrm{wind}}\simeq3\times10^{31}
\left( \frac{\dot{M}}{10^{-6}~\mathrm{M}_{\odot}~\mathrm{yr}^{-1}} \right)
\left( \frac{V_w}{10~\mathrm{km~s}^{-1}} \right)^2~\mathrm{erg~s}^{-1}~~,
\end{equation}
where $\dot{M}$ is the mass-loss rate and $V_w$ is the wind velocity.
An examination of eqs.~(9) and (10) reveals that for the companion
to have an appreciable effect on mass loss (i.e.\ to have 
$\dot{E}_B>0.1\dot{E}_\mathrm{wind}$) it must increase the angular
momentum of the envelope (and thereby $\omega_o/\omega_k$) by an order 
of magnitude (or more). Consequently, in our estimate of the fraction 
of stars that are affected significantly by planetary companions, we 
shall adopt as the defining condition
\begin{equation}
J_p>10J_*~~,
\end{equation}
where $J_p$ is the orbital angular momentum of the planet, and
$J_*$ is the angular momentum of the star as it leaves the main sequence [eq.~(2)]. Note that this is probably a reasonable criterion for the planet having a significant effect on stellar evolution even considering the uncertain nature of the effects on mass loss. In fact, given that the magnetic field in stellar spots can be $\sim10^3$ times stronger than the average field 
\citep[e.g.][]{priest87}, condition~(11) may be too stringent, since even a field lower by an order of magnitude than that given by eq.~(8) will result in spots that are significantly cooler than their surroundings (magnetic pressure larger than the photospheric pressure).

\section{Statistics on Importance of Planets}

We define the following quantities
\begin{eqnarray}
\nu &\equiv&\log (a/AU)\nonumber\\
\mu &\equiv&\log (M/M_J)
\end{eqnarray}
where $a$ is the orbital separation, $M$ is the minimum planetary mass,
and $M_J$ is Jupiter's mass.  In Fig.~1 we show the known 75 planets
(taken in January 2002 from the \textit{Extrasolar Planets Encyclopaedia}
compiled by J.~Schneider, at http://www.obspm.fr/encycl/encycl.html)
in the $\nu$--$\mu$ plane.  In planetary systems where more than one
planet is known we mark the most massive planet by an asterisk, while
the others are marked by the + sign.
Several authors have attempted recently to obtain unbiased
distributions of planets as a function of $\nu$ and $\mu$
\citep[e.g.][]{lineweaver02,mazeh02,tabachnik02,armitage02}.
In view of the uncertainties, we adopt the following simple form for the fraction of systems $dN$ within a box $d\nu d\mu$ (for the  parameter space indicated below)
\begin{equation}
dN=C(1+\alpha\nu)(1-\beta\mu)d\mu d\nu~~,
\end{equation}
for $-1.5<\nu <0.5$ and $\mu<0.5\nu +0.75$. Here $C$, $\alpha$ and $\beta$ are parameters. This form is the same as the one used by 
\citet{lineweaver02},
who found (their figures~3 and~4) $\alpha \simeq 0.8$ and $\beta \simeq 0.8$.
Using the well populated parallelogram 
\citep[i.e.\ excluding areas in the $\nu-\mu$ plane with large gaps; see also][]{zucker02}
marked by lines~1 to~4 on Figure~1, we estimate $\alpha \simeq 0.3$ and $\beta \simeq 0.5$. Examining in addition the results of 
\citet{mazeh02},
\citet{tabachnik02},
and \citet{armitage02},
and fitting where necessary their functional forms to ours, gives average values of $<$$\alpha$$> \simeq0.6$ and $<$$\beta$$> \simeq0.5$, which we shall now adopt.

The value of $C$ can be determined for fixed values of $\alpha$ and $\beta$ in the following way.  From a sample of 1,200 stars discussed by 
Vogt et~al.\ (2002; where more details can be found) 
44~planets and 4~brown dwarfs have been detected. Hence 3.7\% of the surveyed stars were found to have planets.  The 75 known extrasolar planets (as of the end of January 2002) are distributed in 67~planetary systems. Using this ratio we find that the fraction of stars being detected to harbor planetary systems is $\sim 3.3$\%.  If we examine the well populated area of Figure~1 (bounded by the four straight lines $\nu = 0.5$; $\mu = 0.5 \nu +0.75$; $\nu =-1.5$;  and $\mu = 0.5 \nu$), we find that out of the 67~planetary systems, 48 are inside this parallelogram. This corresponds to $\sim 2.4$\% of the stars having their massive planet within this region.  Considering that the right side of this region is close to the detection limit (no planets are found near $\nu \simeq 0.4$ and $\mu \sim 0.3$), we take the detection fraction within this region to be $N_s \simeq 2.5$\% of all stars.  For stars with a metallicity above solar, [Fe/H]~$>0$, the detection fraction is $\sim 2.6$ times higher than that for the entire sample 
\citep{vogt02}. Hence, in the same scaling-region the fraction is $N_s \simeq 6.5$\%. The large increase with metallicity in the probability of harboring a planet  was studied by 
\citet{gonzalez97} and 
\citet{reid02}. The latter author argues that most stars with a metallicity [Fe/H]~$>0.3$ have planets around them.
 We therefore take $N_s=0.025$ for all stars and $N_s=0.065$ for metal rich stars, 
and find the value of $C$ from the expression 
\begin{equation} 
N_s= C \int_{-1.5}^{0.5} d \nu \int_{0.5 \nu}^{0.5 \nu +0.75} 
d \mu   (1+ \alpha \nu)(1- \beta \mu) 
= \frac{C}{32}(48-24 \alpha -6 \beta - 5 \alpha \beta) =
0.9094C.
\end{equation}   
where the last equality was obtained by substituting the average values 
$\alpha=0.6$ and $\beta=0.5$.

In the previous section we derived as our condition for a significant effect on mass-loss from the parent star, $J_p>10J_*$. We noted, however, that it is possible that  even a lower angular momentum might play some role, especially since the star loses angular momentum as it evolves 
\citep{soker01}.
Soker further argues that planets of masses as low as $0.01 M_J$ are sufficient to influence the mass loss \textit{geometry} from AGB stars. An indirect hint that even a relatively slow rotation may influence the mass loss rate comes from the distribution of stars on the horizontal branch in globular clusters (GCs). 
\citet{dcruz96}
show that the distribution of stars on the horizontal branch in the HR diagram can be explained by variations in the mass loss rate on the RGB, by a factor which depends on the metallicity (ranging from 1.5 in metal rich GCs to $\gtrsim 5$ in metal poor GCs). We should note that red horizontal branch stars also rotate \citep{stefanik01}. However, using the results of \citet{behr00} for M13, \citet{soker00} showed that the average angular momentum of the progenitors of hotter horizontal branch stars is larger than that of cooler ones. It is thought that rotation is the main factor that influences the mass loss rate variations (R.~Rood, private communication, 2002), with the exception of very high mass loss rates which may result from interactions with stellar binary companions. Since main sequence stars in GCs are expected to rotate very slowly (since they have experienced a long period of angular momentum losses due to magnetic activity), if the rotation conjecture is correct, then very slow rotation is sufficient to influence mass loss on the RGB. Our condition, therefore, may be somewhat conservative.

The present angular momentum of the Sun is $1.7 \times 10^{48}$g~cm$^2$~s$^{-1}$.
A planet with a Jupiter mass at 1~AU from a solar type star has $\sim 50$ times the present angular momentum of the Sun. Taking as a typical value $J_* = J_\odot$, our condition in equation~(11) reads $(M/M_J)(a/~{\rm AU})^{1/2} \gtrsim 0.2 $. For a more general form $J_p>\eta J_*= \eta J_\odot$, our condition reads
\begin{equation}
\mu + 0.5 \nu \gtrsim  \log (\eta/10)-0.7 \equiv j_m~~.   
\end{equation}
In addition, we limit the planet to $a<2$~AU (or $\nu < 0.3$), in order for tidal forces to be able to bring it into the RGB or AGB star's envelope. These conditions are represented (for $\eta=10$) as the two thick lines in Fig.~1. Planets to the left of and above these lines affect the evolution of their parent stars significantly (neglecting eccentric orbits of planets at larger orbital separations).

Not including the few systems above the line marked `2' in Fig.~1, the total fraction of planetary systems that are likely to significantly influence the evolution of the central star at late stages is therefore, 
\begin{equation}
N({\rm total})= C \int_{-1.5}^{0.3} d \nu \int_{j_m-0.5 \nu}^{0.5 \nu +0.75} 
d \mu   (1+ \alpha \nu)(1- \beta \mu)~~.  
\end{equation}
Note that this integral is meanigful for $j_m<-0.7$. For $j_m>-0.7$ the thick diagonal line in Fig.~1 crosses the upper limit as given by line `2.' Performing this integral for $\alpha=0.6$ and $\beta=0.5$, and substituting for $C$ from equation~(14), we find \begin{equation}
N(total)=N_s(0.4-1.2j_m+0.3 j_m^2)~~.
\end{equation}

Equation~(17) shows that for our strong condition of $\eta =10$, we find that more than 3.5\% of all solar-type stars are significantly affected (in terms of rotation rate and mass loss) by planets or brown dwarf companions.  This number increases to more than 9\% for metal rich (above solar) stars. Relaxing the condition to $\eta=1$, i.e.\ when the planet's orbital angular momentum equals the present one for the sun, increases the numbers to be $>8$\% and $>21$\% respectively. All of these numbers should probably be regarded as lower limits, since we have neglected the effects of planets at 2~AU$\lesssim a\lesssim5$~AU in eccentric orbits 
\citep[which could also be pulled into the giant's envelope by tidal interaction; see also][]{debes02},
and used only the minimum masses (as well as not having included the planets above line `2' in Fig.~1). 

\section{SUMMARY}
We have shown that the swallowing of planetary or brown dwarf companions can have significant effects in the late stages of the evolution of late-type stars.  Specifically, we considered rotation and mass loss. On the basis of the properties of the observed extrasolar planets, we estimated that $\sim4$--10\% of the stars should experience enhanced mass loss due to the effects of planets.

\acknowledgments
We acknowledge useful comments by an anonymous referee.
N.~S.\ was supported by a Celerity Foundation Distinguished Visiting 
Scholar grant at the University of Virginia, and by a grant from the
US-Israel Binational Science Foundation. He also asknowledges the hospitality of ST~ScI.

\clearpage
\begin{figure}
\plotone{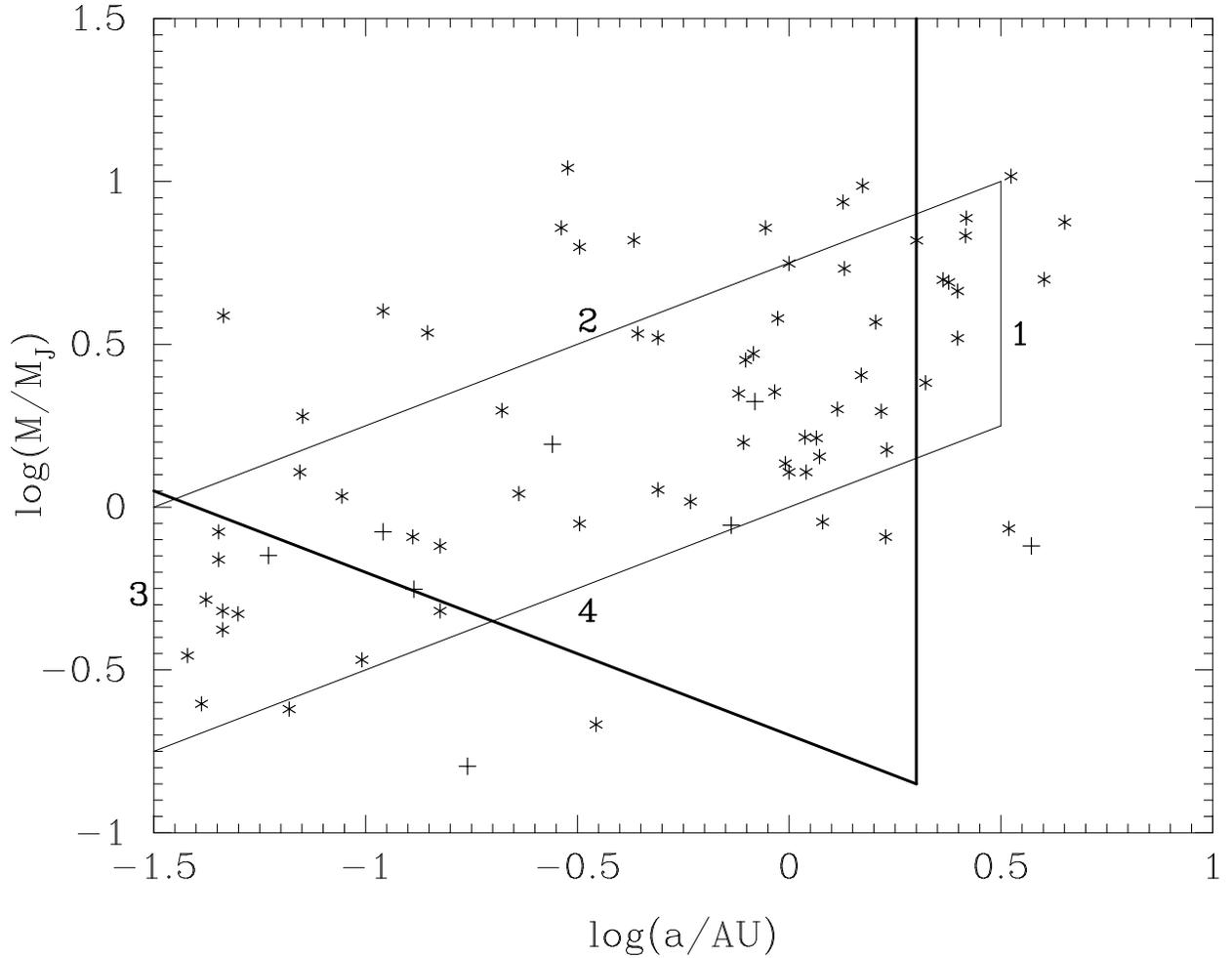}
\caption{75 known extrasolar planets are presented in the minimum-mass--semimajor-axis logarithmic plane.  In planetary systems where more than one massive planet has been found, the most massive one is marked by *, while the others by $+$. The four lines marked~1 to~4 define the area used for our scaling of fraction of planets. The two thick lines represent our condition for a planet to significantly influence the mass loss rate of its parent star (see text). }
\end{figure}
\end{document}